\documentclass[twocolumn,showpacs,preprintnumbers,amsmath,amssymb]{revtex4}

\usepackage{graphicx}
\usepackage{dcolumn}
\usepackage{bm}

\begin{document}


\title{Dynamical evolution of quintessence dark energy in collapsing dark matter halos}

\author{Qiao Wang}
\author{Zuhui Fan}
\affiliation{Department of Astronomy, School of Physics, Peking
University,Beijing, 100871, People Republic of China}

\date{\today}

\begin{abstract}
In this paper, we analyze the dynamical evolution of quintessence
dark energy induced by the collapse of dark matter halos.
Different from other previous studies, we develop a numerical
strategy which allows us to calculate the dark energy evolution
for the entire history of the spherical collapse of dark matter
halos, without the need of separate treatments for linear,
quasi-linear and nonlinear stages of the halo formation. It is
found that the dark energy perturbations evolve with redshifts,
and their specific behaviors depend on the quintessence potential
as well as the collapsing process. The overall energy density
perturbation is at the level of $10^{-6}$ for cluster-sized halos.
The perturbation amplitude decreases with the decrease of the halo
mass. At a given redshift, the dark energy perturbation changes
with the radius to the halo center, and can be either positive or
negative depending on the contrast of $\partial_t \phi$,
$\partial_r \phi$ and $\phi$ with respect to the background, where
$\phi$ is the quintessence field. For shells where the contrast of
$\partial_r \phi$ is dominant, the dark energy perturbation is
positive and can be as high as about $10^{-5}$.
\end{abstract}

\maketitle

\section{Introduction}
The fast advance of different cosmological observations permits us
to weigh the universe quantitatively \cite{Astier06,Riess07}. It
is known that the universe is currently dominated by the dark
energy component, which accounts for about $70\%$ of the total
energy density. Probing the nature of dark energy is thus one of
the most challenging tasks in cosmological studies. The simplest
dark energy candidate is the cosmological constant with the
equation of state $w=-1$. While being consistent with
observations, it runs into severe difficulties theoretically,
namely the fine tuning problem and the coincidence problem.
Various dynamical dark energy models have been proposed aiming to
avoid such problems \cite{Copeland06,Frieman08,Linder08}. The
allowed parameter space for dynamical dark energy models derived
from current observational data is still relatively large
\cite{Komatsu09,Li08JCAP}. Tighter constraints are highly expected
from future observational data with much improved quality and
quantity. To realize the goal, however, it is crucial to
understand in detail the observational impacts from the dark
energy component. Different dark energy models give rise to
different expansion histories of the universe, which in turn
influence differently on the formation and evolution of
large-scale structures
\cite{Klypin03,Maccio05,LeDelliou06,Mainini07}. Besides,
clustering can occur for dynamical dark energies themselves, and
the clustering characteristics are highly model dependent.
Therefore the observable imprints from dark energy perturbations
can potentially affect the constraints on the properties of dark
energy.

Dark energy perturbations associated with the linear evolution of
the matter inhomogeneities have been studied extensively
\cite{Unnikr08,Amendola04}. Considerable effects can occur. For
instance, it has been found that with the proper inclusion of dark
energy perturbations, the constraints on the time-evolving
equation of state of dark energy are weakened significantly due to
their influence on the late ISW effects of CMB anisotropy
\cite{Zhao05,Zhao07,Li08ApJ}. On the other hand, the dark energy
clustering related to the nonlinear structure formation is less
addressed. This is partially due to the numerical difficulties in
solving the coupled perturbation equations for matter and dark
energy in the nonlinear regime. However, many cosmological
observations, such as the abundance of clusters of galaxies and
weak gravitational lensing effects on small scales, are directly
related to the nonlinear structure formation \cite{LaVacca08}. It
has been proposed that the recently reported cold spots seen on
the CMB map may arise from the Rees-Sciama effect, the nonlinear
version of the ISW effects \cite{Granett08,Sakai08,Masina09}.
Therefore there is a need in understanding thoroughly the effects
of dark energy on the formation of nonlinear structures. It is
also of interest to investigate the corresponding clustering
behavior of dark energy.

There have been analyses on the formation of nonlinear structures,
such as the virial radius and the average mass density of halos,
taking into account the existence of the background dark energy
\cite{WS98,ML04,MB04,NM06}. Parameterized dark energy
perturbations and their correlations with the structure formation
are investigated in some of the studies
\cite{Langlois07,Abramo07,ML04, MOTA08JCAP}. Recently, under the
framework of spherical collapse of halos, the perturbations of
quintessence fields induced by the structure formation are
analyzed \cite{DM07,MSS08}. Different from the parameterization
approach, in which the equation of state and the perturbation of
dark energy are regarded as being independent with each other,
such analyses can present a consistent picture on the dark energy
perturbation that often correlates with the evolution of the
equation of state strongly. Dutta and Maor \cite{DM07} mainly
study the dark energy perturbation in the linear regime of the
halo formation. Mota, Shaw $\&$ Silk \cite{MSS08} consider the
dark energy clustering in the process of nonlinear structure
formation employing a matched asymptotic expansion analysis
method, in which, they adopt different parameter expansions for
the linear, quasi -linear and nonlinear stages, respectively. In
this paper, we present a numerical method within the spherical
collapse model that allows us to calculate the co-evolution of the
halo formation and the dark energy clustering during the whole
process, from linear to nonlinear epoches. In this approach, we do
not need to treat the linear, quasi-linear and nonlinear
evolutions separately. We analyze the quintessence dark energy
perturbation paying particular attention to its correlation with
the background field evolution, i.e., the evolution of the
equation of state of dark energy.

The paper is organized as follows. In section 2, we present the
model and the corresponding formulae. Section 3 explains in detail
our numerical approach. Our results are shown in section 4.
Section 5 contains discussions.

\section{The Model and Formulae}
\subsection{Two-components spherical collapse model}
In the framework of the spherical collapse of dark matter halos,
we consider the induced perturbation of dark energy. We start with
the line element of a spherically symmetric system, which can be
written as
\begin{equation}
ds^2 = -dt^2+e^\mu dr^2+R^2(d\theta^2+\sin^2 \theta d\varphi^2)
\end{equation}
where $\mu$ and R are functions of both time $t$ and radius $r$.
Note the corresponding $R=a(t)r$ in the Friedmann-Robertson-Walker
(FRW) metric, where $a(t)$ is the scale factor of the universe.
The coordinate system is chosen such that it is comoving with the
dark matter component. Thus the $tt$ component of the
energy-momentum tensor for dark matter is equal to the matter
density, and the other components are zero. For the quintessence
model, the energy-momentum tensor is
\begin{equation}
T_{\alpha \beta}^{\phi}=\partial _\alpha \phi \partial _\beta \phi
- g_{\alpha \beta}L ,
\end{equation}
and
\begin{equation}
L=\frac{1}{2}\partial^\alpha \phi \partial _\alpha \phi + V(\phi),
\end{equation}
where V($\phi$) is the potential of the quintessence. The total
energy-momentum $T_{\mu \nu}$ can therefore be expressed as
\begin{subequations}
\begin{equation}
T_{tt}=\rho _m  + \frac{1}{2}\dot \phi ^2  + \frac{{e^{ - \mu }
}}{2} \phi '^2 + V\left( \phi  \right)  ,
\end{equation}
\begin{equation}
T_{rr}=e^\mu \left(\frac{1}{2}\dot \phi ^2 +  \frac{{e^{ - \mu }
}}{2} \phi '^2 - V\left( \phi  \right) \right) ,
\end{equation}
\begin{equation}
T_{\theta \theta}=R^2\left( \frac{1}{2}\dot \phi ^2 - \frac{{e^{ -
\mu } }}{2}\phi '^2  - V\left( \phi
   \right)\right),
\end{equation}
\begin{equation}
T_{\varphi \varphi} = \sin ^2 \theta T_{\theta \theta},
\end{equation}
\begin{equation}
T_{tr}=\dot \phi \phi ',
\end{equation}
\end{subequations}
where the dot denotes the time derivative $ {\partial}/{\partial
t}$ and the prime denotes the spatial derivative $
{\partial}/{\partial r}$.

From the Einstein field equations, we have
\begin{subequations}

\begin{eqnarray}
- \frac{{2\ddot R}}{R} + \frac{{\dot \mu \dot R}}{R} &+&
\frac{{e^{- \mu } }}{R}\left( {\mu 'R' - 2R''} \right) \nonumber\\
 & &= 8\pi G\left({\rho _m  + \dot \phi ^2  + e^{ - \mu } \phi '^2 }
 \right),
\end{eqnarray}
\begin{eqnarray}
\frac{{2\ddot R}}{R} + \frac{{\dot \mu \dot R}}{R} + \frac{{2\dot
R^2 }}{{R^2 }} &+& \frac{2}{{R^2 }} + e^{ - \mu } \left(
{\frac{{\mu 'R'}}{R} - \frac{{2R''}}{R}
- \frac{{2R'^2 }}{{R^2 }}} \right) \nonumber \\
 & &=8\pi G\left( {\rho _m  + 2V} \right),
\end{eqnarray}
\begin{equation}
\frac{{\dot \mu R'}}{R} - \frac{{2\dot R'}}{R} = 8\pi G\dot \phi
\phi '. \label{eq:dp}
\end{equation}
\end{subequations}

The evolution of the quintessence scalar field, which is not
coupled to the dark matter component except gravitationally, is
governed by
\begin{equation}
\frac{1}{{\sqrt {-g} }}\partial _\alpha  \left( {\sqrt {-g
}g^{\alpha \beta }
\partial _\beta  \phi } \right) + \frac{{dV}}{{d\phi }} = 0,
\label{eq:KG}
\end{equation}
where $-g=\hbox{Det}(g_{\alpha\beta})$.

\subsection{Tolman approximation}
The perturbation of non-coupled quintessence dark energy is
expected to be small. Thus to a good approximation, we can ignore
the time-spatial term $\dot \phi \phi '$ in the right hand side of
the equation (\ref{eq:dp}). By time integrating its left hand
side, we obtain a relation between $\mu $ and R, which follows
\begin{equation}
e^\mu   = \frac{{R'^2 }}{{1 + f\left( r \right)}}, \label{eq:muR}
\end{equation}
where $f$ is a constant of integration and a function of $r$ only,
which reflects the local spatial curvature.  Then we have
\begin{equation}
ds^2  =  - dt^2  + \frac{{R'^2 }}{{1 + f\left( r \right)}}dr^2  +
R^2 \left( {d\theta ^2  + \sin ^2 \theta d\varphi ^2 } \right).
\end{equation}
This is the well-known Tolman metric that is often utilized to
compute the evolution of spherical objects. For a homogeneous
universe, the Tolman metric reduces to the FRW metric and $f\left(
r \right)=-kr^2$. With the Tolman metric, the Einstein equations
become
\begin{subequations}
\begin{equation}
\frac{{\ddot R}}{R} + \frac{1}{2}\left( {\frac{{\dot R^2 }}{{R^2
}} - \frac{f}{{R^2 }}} \right) =  - 4\pi G\left( {\frac{1}{2}\dot
\phi ^2  + \frac{{e^{ - \mu } }}{2}\phi '^2  - V} \right),
\label{eq:TolEina}
\end{equation}
and
\begin{equation}
\frac{{\dot R^2 }}{{R^2 }} - \frac{f}{{R^2 }} + \frac{{2\dot R\dot
R'}}{{RR'}} - \frac{{f'}}{{RR'}} = 8\pi G\left( {\rho _m  +
\frac{1}{2}\dot \phi ^2  + \frac{{e^{ - \mu } }}{2}\phi '^2  + V}
\right). \label{eq:TolEinb}
\end{equation}
\end{subequations}

It is noted that keeping the $\phi '^2$ term formally does not
affect our derivation of equations (\ref{eq:TolEina}) and
(\ref{eq:TolEinb}). On the other hand, to be consistent, we do not
include it when numerically calculating the evolution of the
metric $R$.

The dynamical equation of quintessence (\ref{eq:KG}) in the Tolman
metric is then written as
\begin{eqnarray}
\label{eq:Tolsf} \ddot \phi  &+& \left( {\frac{{\dot R'}}{{R'}} +
\frac{{2\dot R}}{R}} \right)\dot \phi + \frac{{dV}}{{d\phi }} \\
& & - \frac{{1 + f}}{{R'^2 }}\left[ {\phi '' + \left(
{\frac{{2R'}}{R} + \frac{{f'}}{{2\left( {1 + f} \right)}} -
\frac{{R''}}{{R'}}} \right)\phi '} \right] = 0.  \nonumber
\end{eqnarray}
The equations (\ref{eq:TolEina}), (\ref{eq:TolEinb}) and the
equation (\ref{eq:Tolsf}) constitute the basic dynamical equations
in our analyses.

\subsection{Spatial Curvature $f(r)$}
In the Tolman metric approximation, the term $f(r)$ does not
depend on time. Therefore we can calculate this quantity at any
chosen time. In our consideration, we calculate $f(r)$ at the
initial redshift chosen to be $z_i = 30$. Because the dark energy
component is negligible at such a high redshift, the equations
(\ref{eq:TolEina}) and (\ref{eq:TolEinb}) can be further
simplified. By combining the two equations, we obtain
\begin{equation}
 f(r) =  \dot R^2|_{z_i} - \bigg [\frac{{2GM\left( r \right)}}{R}\bigg ]_{z_i},
\label{eq:ff}
\end{equation}
where $M\left( r \right)$ denotes the mass enclosed within the
comoving radius r, and
\begin{equation}
\frac{dM\left( r \right)}{dr}|_{z_i} = 4\pi \rho \left( {z_i ,r}
\right)(R^2 R')|_{z_i}.
\label{eq:mass}
\end{equation}
Therefore $f\left( r \right)$ depends on the initial dark matter
density profile. In this paper, we adopt the following density
distribution
\begin{equation}
\rho \left( {t_i ,r} \right)=\rho_m\left({t_i}\right)\left[
{1+\delta _c e^{-\frac{{r^2 }}{{\sigma^2
}}}}\left(1-\frac{2r^2}{3\sigma^2}\right) \right]
\label{eq:profile}
\end{equation}
where $\rho_m\left({t_i}\right)$ is the initial background dark
matter density, $\delta _c$ is the initial central density
contrast in the halo region, and $\sigma$ is related to the
characteristic scale of the halo and will be discussed further
later. For $\delta _c>0$/$\delta _c<0$, such a perturbation
profile has an overdense/underdense zone in the central region
and an underdense/overdense transitional middle region which
compensates for the central region so that the overall
perturbation is zero.

\subsection{The double exponential quintessence model}

In this paper, we consider the quintessence model with a commonly
adopted double exponential potential
\begin{equation}\label{21}
V\left( \phi  \right) = V_0 \left( {e^{\alpha \sqrt {8\pi G} \phi
}  + e^{\beta \sqrt {8\pi G} \phi } } \right),
\end{equation}
where $\alpha$ and $\beta$ are parameters representing the
profiles of the two exponentials, respectively, and $V_0$ reflects
the height of the potential. In our analyses, we take $\alpha=6$
and $\beta=0.1$ as our fiducial model. We also vary $\alpha$ with
$\alpha=5, \hbox{ } 6, \hbox{ } 7$ to see the effects of the
potential on dark energy perturbations. The starting redshift is
taken to be $z_i=30$. The initial velocity of the $\phi$ field is
taken to be $\dot \phi=0$. The $V_0$ parameter and the initial
value of $\phi$ are adjusted so that we have $\Omega_{DE}= 0.7$
and $\omega\approx -0.95 $ at $z=0$. Specifically, we first choose
a pair of reasonable values of $V_0$ and the initial $\phi$, and
evolve the background field to the present to calculate
$\Omega_{DE}$ and $\omega$ at $z=0$. By comparing the obtained
$\Omega_{DE}$ and $\omega$ with $\Omega_{DE}= 0.7$ and
$\omega\approx -0.95$, we adjust the values of $V_0$ and the
initial $\phi$ to compute the evolution of the field again. Such a
process is repeated until $\Omega_{DE}= 0.7$ and $\omega\approx
-0.95$ are reached. It is known that the double exponential
quintessence model has a late-time attractor solution, in which
the final $\Omega_{DE}$ and $\omega$ are mostly determined by the
potential, i.e., the values of $V_0$, $\alpha$ and $\beta$, but
not very sensitive to the initial conditions of $\phi$
\cite{Barreiro00}. Indeed, such a tracking behavior has been seen
in our analyses. On the other hand, we start from a relatively low
redshift ($z_i=30$), and choose $\dot \phi=0$ initially. Thus
albeit not an extremely fine-tuning process, a certain level of
adjustment on the initial value of $\phi$ is needed in our
analyses.

Including the perturbations, we have
\begin{subequations}
\begin{equation}
\rho = \frac{\dot \phi ^2}{2}  + \frac{\left( {\nabla \phi }
\right)}{2}^2 + V\left( \phi \right) \label{eq:sfrho},
\end{equation}
\begin{equation}
p = \frac{\dot \phi ^2}{2}  - \frac{\left( {\nabla \phi }
\right)^2}{6}  - V\left( \phi  \right) \label{eq:sfprs},
\end{equation}
and
\begin{equation}
\omega  = \frac{p}{\rho }\quad .
\label{eq:sfeos}
\end{equation}
\end{subequations}

    \begin{figure}
    \includegraphics[width=0.42\textwidth]{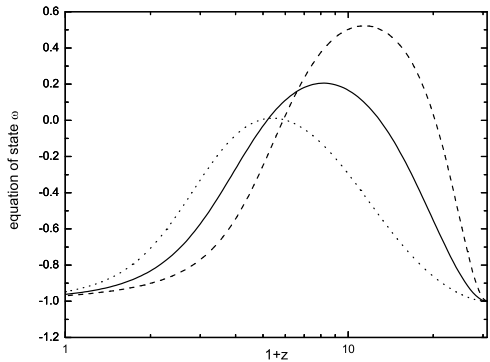}
    \caption{\label{fig:eos57} The evolution of the background equation of state for the
    double exponential quintessence models. We take $\beta =0.1$, and the initial
    redshift $z_i = 30$. The dotted, solid and dashed lines are
    for $\alpha = 5, 6$ and $7$, respectively.}
    \end{figure}

    \begin{figure}
    \includegraphics[width=0.42\textwidth]{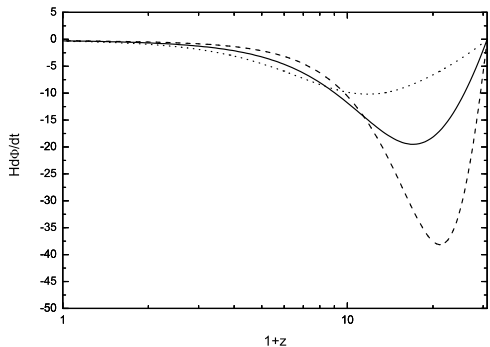}
    \caption{\label{fig:psi57}The evolution of the background $H\dot \phi$. The parameters
    are the same as in Fig.~\ref{fig:eos57}.}
    \end{figure}

In Fig.~\ref{fig:eos57}, we show the evolution of the background
equation of state $\omega$ with $\nabla \phi = 0 $ for $\alpha=5,
\hbox{ } 6,\hbox{ and } 7$, respectively. The corresponding
evolution of the time derivative of the background quintessence
field, $H\dot \phi$, are shown in Fig.~\ref{fig:psi57}. It is seen
that with the increase of $\alpha$, the potential gets steeper and
the $H\dot \phi$ term is larger. As a result, the equation of
state $\omega$ increases. For $\alpha=6$ and $\alpha=7$,
$\omega>0$ in the redshift ranges $5<z<14$ and $6<z<20$,
respectively. It will be seen later that the evolutionary behavior
of the background field affects the dark energy perturbations
significantly, thus in general, the two cannot be treated
independently.

\section{Numerical method}
\subsection{Iterative algorithm}
We first compute the local metric in the region of the spherical
dark matter halo. We incorporate the local dark matter density and
the dynamical evolution of the background quintessence field into
our calculation. The dark energy perturbations are neglected in
the metric calculations. Thus equation(\ref{eq:TolEina}) can be
written as follows
\begin{equation}
\frac{{\ddot R}}{R} + \frac{1}{2}\left( {\frac{{\dot R^2 }}{{R^2
}} - \frac{f}{{R^2 }}} \right) =  - 4\pi G\left( {\frac{1}{2}\dot
\phi_{bg} ^2  - V\left(\phi_{bg} \right)} \right)
\label{eq:NumEin}
\end{equation}
where $bg$ stands for background. Equation (\ref{eq:TolEinb}) has
been used in deriving the expression of $f$ given in part C of
the previous section.

Equation (\ref{eq:NumEin}) is an ordinary differential equation.
We then solve it at different shells equally spaced in $r$. The initial
conditions for $R$ are chosen to be $R(z_i)=a(z_i)r$ and $\dot R=\dot a(z_i)r$ at $z_i=30$,
where $a(z_i)$ is the scale factor of the universe at $z_i$.

With the obtained metric, we thus can analyze the spatial
perturbation of dark energy. We rewrite the
equation(\ref{eq:Tolsf}) in the following form

\begin{eqnarray}
\label{eq:Numsf} \phi '' &+& \left( {\frac{{2R'}}{R} +
\frac{{f'}}{{2\left( {1 + f}
\right)}} - \frac{{R''}}{{R'}}} \right)\phi ' \nonumber \\
& &= \frac{{R'^2 }}{{1 + f}}\left( {\ddot \phi  + \left(
{\frac{{\dot R'}}{{R'}} + \frac{{2\dot R}}{R}} \right)\dot \phi  +
\frac{{d V}}{{d\phi }}} \right).
\end{eqnarray}

Equation (\ref{eq:Numsf}) can be regarded as a differential
equation of $\phi$ with respect to $r$ with the right hand side as
the source of the spatial perturbations of $\phi$. We solve this
equation iteratively at each time $t$. First, we insert the
background $\phi_{bg}(t)$ and $\dot \phi_{bg}(t)$ into the right
hand side of equation (\ref{eq:Numsf}). With this source term, we
solve $\phi (r,t)$ at each $r$. Then for each obtained $\phi
(r,t)$, we find its corresponding $\dot \phi_{bg}$. This is done
by looking through the evolution of the background field $\phi$,
and locating its $\dot \phi_{bg}$ value at the place where
$\phi=\phi (r,t)$. Then at this time, different $\dot \phi_{bg}$
value is obtained at different $r$. In the next iteration, this
set of $\dot \phi_{bg}(r,t)$ are used in calculating the source
term, which lead to a new set of $\phi (r,t)$. We repeat such an
iteration process until the solutions on $\phi (r,t)$ converge.
For each iteration, the shooting method from the center outward
is applied to solve this spatial differential
equation \cite{NR02}. The boundary conditions are taken to be $\phi
\left( r_{bg} \right) = \phi_{bg}$ and $
\partial \phi/\partial r|_{r=0} = 0$. As shown in Eq. (\ref{eq:profile}),
we consider compensated matter density perturbations. Thus $r_{bg}$ is defined to be
the shell inside which the average matter density perturbation is close to zero. The evolution
of the shell at $r_{bg}$ follows that of the background universe.

In comparison with the method solving the partial differential equation
directly, our numerical algorithm not only saves time but
also avoids propagating numerical errors step by step in the time
advancement.

\subsection{Initial conditions}

In this study, we consider flat cosmologies with the parameters
being taken from WMAP5 results \cite{WMAP5}. Specifically,
$\Omega_m = 0.249$, $h = 0.72$ and $\sigma_8 = 0.787$, where
$\Omega_m$ is the dimensionless matter density of the universe,
$h$ is the Hubble constant in units of $100\hbox{ km/s/Mpc}$, and
$\sigma_8$ is the linearly extrapolated $rms$ matter density
fluctuation on top-hat scale of $8\hbox{ Mpc}h^{-1}$.
The initial redshift is taken to be $z_i=30$.
We consider overdense regions with typical scales of clusters and
galaxies, respectively. For galaxies, we take $M=10^{12}
M_{\odot}$, and for clusters of galaxies, we take $M=10^{14}
M_{\odot}$. The corresponding linear comoving scales are calculated
through $M=(4\pi/3)\Omega_m\rho_c r_*^3$, where $\rho_c$ is the critical
density of the universe at present. This gives rise to $r_*=1.8 \hbox{
Mpc}$ and $r_*=8.4 \hbox{ Mpc}$, respectively. The initial
perturbations over the considered scales are respectively set to
be $\bar \delta_i=0.11$ and $\bar \delta_i=0.044$, estimated by
employing the linear power spectrum of Eisenstein \& Hu
\cite{EH98}.

We now derive the relation between $\delta_c$ and $\sigma$ in
equation (\ref{eq:profile}) and $\bar \delta_i$ and $r_*$. From
the definition of $\bar \delta_i$, we have

\begin{equation}
\int_0^{r_*} \delta_i(r) 4\pi r^2 dr = \frac{4\pi}{3}\bar \delta_i
r_*^3,
 \nonumber
\end{equation}
where
\begin{equation}
\delta_i(r)= \delta _c e^{-\frac{{r^2 }}{{\sigma^2
}}}\left(1-\frac{2r^2}{3\sigma^2}\right).
\label{eq:delta}
\end{equation}
Then we obtain
\begin{equation}
\delta _c e^{-\frac{{r_*^2 }}{{\sigma^2 }}}=\bar \delta_i.
 \nonumber
\end{equation}
In our analyses, we take $\delta_c=5\bar \delta_i$, therefore we
have $\sigma=r_*/1.27$. The largest radius in our calculation is
set to be $r_{bg}=6.5\hbox{ Mpc}$ for the galaxy scale and $r_{bg}=30\hbox{ Mpc}$
for the cluster scale. Beyond the largest radius, all the relevant
quantities merge into the background ones.

For the curvature term $f \left( r \right)$, as we discussed
previously, it can be calculated at $z_i$. Thus we have
\begin{equation}
f = H ^2 \left({z_i} \right)R_i ^2-\frac{{2GM\left( r
\right)}}{R_i}.
\end{equation}
We normalize $R$ with the largest scales in our calculations.

    \begin{figure}
    \includegraphics[width=0.45\textwidth]{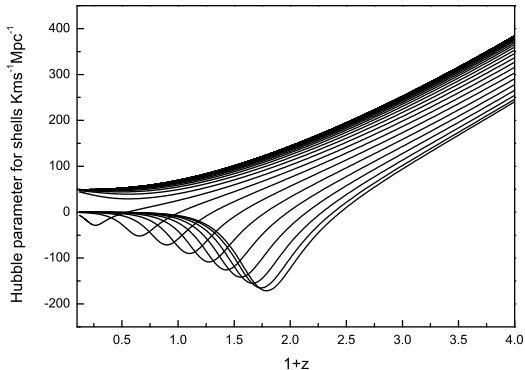}
    \caption{\label{fig:Hubflux}The local Hubble parameter $\dot
    R/R$ for different shells of dark matter. From upper to lower curves,
    $r$ is decreasing.}
    \end{figure}

\subsection{Virialization}
In the general spherical collapse model, the virialization process
is not naturally included. It is known that the violent
realization and phase mixing play important roles in the final
stage of the formation of dark matter halos. However, detailed
understanding and modelling of the virialization process are still
lacking. In \cite{SHAW08,MSS08}, they fit the simulation results
to describe the virialization stage.

Here we adopt the following description for the virialization. The
standard spherical model is used to calculate the metric evolution
of $R$ for each shell before its turn-around, denoted as
$R_{max}$. After the turn-around, the time derivative of $R$ is
modified such that it is approaching zero at the virialization
radius $R_{vir}$. Specifically, we use
\begin{equation}
\label{eq:mod} \frac{\dot R}{\dot R_{SCM}} = \frac{1}{2}\left(
cos(\frac{R - R_{max}}{R_{max} - R_{vir}} \pi) + 1 \right),
\end{equation}
where the $SCM$ stands for the standard spherical collapse model.
It is noted that in our modelling, the overdense region has a
r-dependent density profile. Consequently, different shells reach
turn-around at different time. Thus Eq. (\ref{eq:mod}) models
the virialization for regions inside individual shells. In other
words, the virialization starts from the inner region and
gradually extends outward. To certain extents, we believe that our
modelling grabs the basic picture of the virialization.

Fig.~\ref{fig:Hubflux} shows $\dot R/R$ for some shells, where we
take $R_{vir}(r)=R_{max}(r)/1.8$ instead of 2.0 from simulation results
used in \cite{MSS08,HAMILTON91}. We also calculate the
mean overdensity $\Delta_{vir} = \bar \rho(r) / \rho_{bg} |
_{z_{vir}} $, where $\bar \rho(r)=M(r)/(4\pi R_{vir}(r)^3/ 3)$ is the mean mass
density within $r$ and the mass $M(r)$ contained inside $r$ is computed
by Eq. (\ref{eq:mass}),
and $\rho_{bg}$ is the background density at the virialization
time. Here the virialization redshift $z_{vir}$ for a shell is defined
to be the redshift when the shell collapses to the center in the case without
considering the virialization treatment. For different shells, the results are shown in
Table~\ref{tab:Dltvir}. It is seen that for regions where the
virialization happens at high redshift, $\Delta_{vir}\sim 179$,
the well known value for $\Omega_m=1$. At later times, as the
contribution from the dark energy component becomes significant,
$\Delta_{vir}$ decreases. At $z\sim 0$, we have $\Delta_{vir}\sim
150$. Our results are consistent with those from the simple
consideration of energy partitions \cite{MOTA08JCAP}, indicating
the meaningfulness of our treatment for the virialization.

    \begin{table}
    \caption{\label{tab:Dltvir}Halo does not have a single collapse time in
    our model. This table shows $\Delta_{vir}$ of some shells at the time of their
    collapse. }
    \begin{ruledtabular}
    \begin{tabular}{cddddd}
         $z_{vir}$    & 0.003 &  1.07& 1.99& 3.57 & 4.92\\
    \hline
     $\Delta_{vir}$    &148.96&153.92& 162.33& 172.82& 179.21\\
    \end{tabular}
    \end{ruledtabular}
    \end{table}

    \begin{figure}
    \includegraphics[width=0.45\textwidth]{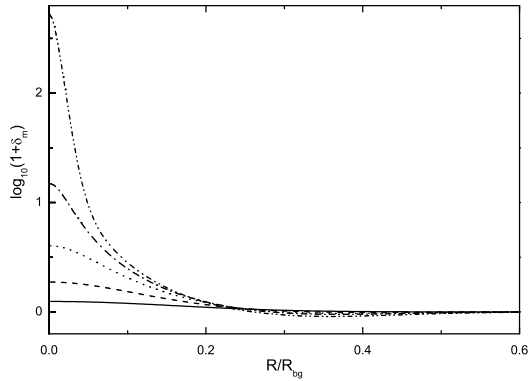}
    \caption{\label{fig:mprofile}The relative mass density profiles at different redshifts.
    The solid, long dashed, dotted, dash-dotted and dot-dot-dashed lines are for
    the results at $z=20, 5, 2, 1,$ and $0$, respectively}
    \end{figure}

In Fig.~\ref{fig:mprofile}, we show the time evolution of the
relative density profile, where the horizontal axis is $R/R_{bg}$
with $R_{bg}$ being the $R$ value at $r_{bg}$ and the verticle
axis is the relative density $\rho(R,z)/\rho_{bg}(z)$. As
expected, the relative density gets higher and more centrally
concentrated as time evolves. On the other hand, it is noted that
our prescription of the virialization Eq. (\ref{eq:mod}) only
accounts for the fact that $R_{vir}/R_{max}\approx 1.8$ for a
virialized shell. It does not include considerations about the
detailed virialization processes, such as shell crossing, phase
mixing and violent relaxation. Thus in our treatment, the
evolution of $R$ is basically an initial value problem, and the
final relative density profile is closely related to its initial
one. Since the NFW profile found from numerical simulations
\cite{NFW96}\cite{NFW97} must be at least partially related to the
virialization processes, we do not expect that the NFW profile can
be naturally obtained in our simple analyses presented here
without adjusting the initial relative density profile carefully.

\section{Result}

In this section, we present the results on the calculated dark
energy perturbations.

\begin{figure}
    \includegraphics[width=0.45\textwidth]{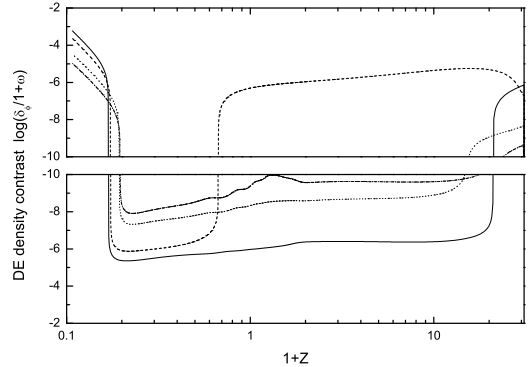}
    \caption{\label{fig:ssss}The logarithmic plot of the evolution of DE density
    contrast of four shells. The upper panel shows the positive part of
    $\delta_{\phi}/(1+\omega)$ and the lower panel is for the negative part.
    The solid, dashed, dotted and dash-dotted lines are for $r=0, 0.25, 0.5$ and $0.75$ respectively.}
    \end{figure}

\begin{figure}
    \includegraphics[width=0.45\textwidth]{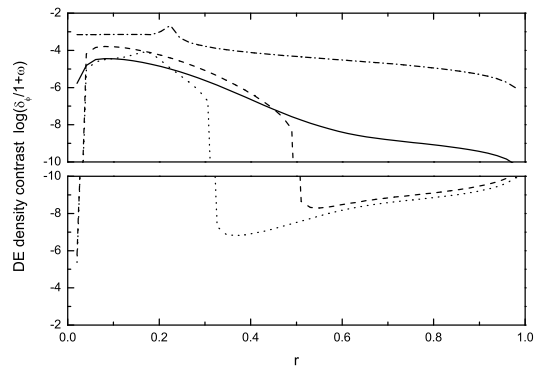}
    \caption{\label{fig:space}The logarithmic plot of the spatial distribution of
    $\delta_{\phi}/(1+\omega)$ at of different redshifts. The upper panel is for
    the positive part of $\delta_{\phi}/(1+\omega)$ and the lower panel is
    for the negative part. The solid, dashed, dotted and dash-dotted lines are for
    $1+z = 22, 10, 1$ and $0.12$, respectively.}
    \end{figure}

Fig.~\ref{fig:ssss} shows the time evolution of
$\delta_{\phi}/(1+\omega)$ for four spatial shells with $r=0$,
$0.25$, $0.5$ and $0.75$, respectively, for the cluster-scale
halo, where $\delta_{\phi}$ is the dark energy density
perturbation, and $r$ is in unit of the largest radius
($30.3\hbox{ Mpc}$ in this case) in our calculation. The
corresponding spatial variations of $\delta_{\phi}/(1+\omega)$ at
different redshifts are shown in Fig.~\ref{fig:space}.
For the purpose of easy comparison with the results of \cite{MSS08},
we follow them to show $\delta_{\phi}/(1+\omega)$ instead of $\delta_{\phi}$.
Under certain approximations, it is shown in \cite{MSS08} that the
quantity $\delta_{\phi}/(1+\omega)$ can be directly related to
the matter density perturbations. This motivates them to focus on
$\delta_{\phi}/(1+\omega)$. In our following presentations,
besides $\delta_{\phi}/(1+\omega)$, we also show $\delta_{\phi}$ at various places.
We can see that the amplitude of $\delta_{\phi}/(1+\omega)$ is
$\sim 10^{-5} -10^{-6}$ except for the central region where
$\delta_{\phi}/(1+\omega)$ can reach as high as about $10^{-4}$.
At $1+z\ge 20$, the dark energy perturbations are positive
everywhere in the calculated region, showing an initial
concentration of dark energy in the region. With the evolution of
the halo formation, the outer region appears as a dark energy void
with negative $\delta_{\phi}/(1+\omega)$, while the inner region
(except for $r=0$) has a positive dark energy perturbation. As
time proceeds, the void region expands till very late time in the
future when the dark energy perturbation turns to be positive
again everywhere in the region.

\begin{figure}
\includegraphics[width=0.47\textwidth]{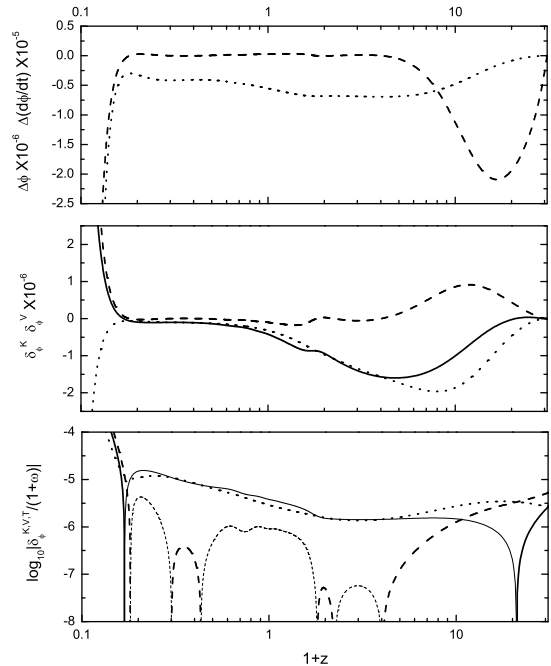}
\caption{\label{fig:sfsfdot1}The evolution of the central density
contrast. Upper panel: The evolution of $\Delta \phi$ (dotted) and
$\Delta \dot \phi$ (dashed); Middle panel: The evolution of
$\delta_{\phi}^K \times 10^{-6}$ (dashed) and $\delta_{\phi} ^V
\times 10^{-6}$ (dotted) and $(\delta_{\phi}^K + \delta_{\phi} ^V)
\times 10^{-6}$ (solid); Lower panel: The evolution of
$log_{10}|\delta_{\phi} ^K/(1+\omega)|$ (dashed) and
$log_{10}|\delta_{\phi} ^V/(1+\omega)|$ (dotted) and
$log_{10}|(\delta_{\phi} ^K + \delta_{\phi} ^V)/(1+\omega)|$
(solid).}
\end{figure}

\begin{figure}
\includegraphics[width=0.45\textwidth]{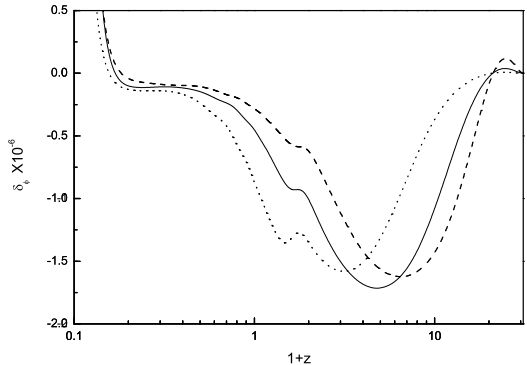}
\caption{\label{fig:alpha} The evolution of $\delta _{\phi}\times
10^{-6}$ for $r=0$. The dotted, solid and dashed lines are for
$\alpha = 5, 6$ and $7$, respectively.}
\end{figure}

In the following, we will analyze different contributions to the
dark energy density perturbations in detail. Generally, there are
three terms to determine the total energy density of a scalar
field, namely, the kinetic energy term $\sim \dot \phi^2$, the
potential energy term $\sim V(\phi)$ and the term related to the
spatial variation of the field $\sim \phi \prime ^2$. For the very
central shell with $r=0$, we choose a boundary condition with
$\phi \prime=0$, and thus for this shell, only the kinetic and the
potential terms exist. In Fig.~\ref{fig:sfsfdot1}, the upper panel
shows the perturbations of the field $\phi$ (dotted line) and its
time derivative $\dot \phi$ (dashed line) with respect to the
background quantities. Because of its slower expansion rate $\dot
R$ than the background, the dragging term $(\dot R/R)\dot \phi$ is
smaller in equation (\ref{eq:Tolsf}). Therefore the evolution of
the field is faster, which leads to smaller value of $\phi$ and
the corresponding change of $\dot \phi$ comparing to those of the
background quantities. The middle panel shows the corresponding
energy perturbations. The lower panel presents
$\delta^K_{\phi}/(1+\omega)$, $\delta^V_{\phi}/(1+\omega)$ and
$(\delta^K_{\phi}+\delta^V_{\phi})/(1+\omega)$ in logarithm scale,
where the thick and thin lines correspond to positive and negative
quantities, respectively. It is seen that the potential energy
perturbation is always negative, while the kinetic energy
perturbation can be positive or negative depending on the shape of
the potential. At early times with $z>20$, the kinetic energy
perturbation is dominant, and thus we have $\delta_{\phi}>0$. At
later times, the potential energy perturbation becomes the
dominant term, which gives rise to the negative dark energy
perturbation, i.e., the dark energy void. It is noted that as the
shell turns over, the term $(\dot R/R)\dot \phi$ changes sign, and
thus it is no longer the friction term, but a driving force for
the local evolution of the field. It causes the deviation of $\dot
\phi$ from the background, and eventually leads to the domination
of the kinetic energy perturbation and thus positive dark energy
perturbations again at very late time with $z<0$. It can be seen
that the behavior of the dark energy perturbations is closely
related to the form of the potential of the field, which in turn
determines the evolution of equation of state of the background
dark energy. Fig.~\ref{fig:alpha} shows the time evolution of the
dark energy perturbations of the central shell for different
potentials with $\alpha=5$ (dotted line), $6$ (solid line) and $7$
(dashed line), respectively. The dependence of $\delta_{\phi}$ on
$\alpha$ is clearly seen. Therefore the dark energy perturbations
and its background E.o.S are in general highly correlated, and
cannot be treated independently.

\begin{figure}
\includegraphics[width=0.47\textwidth]{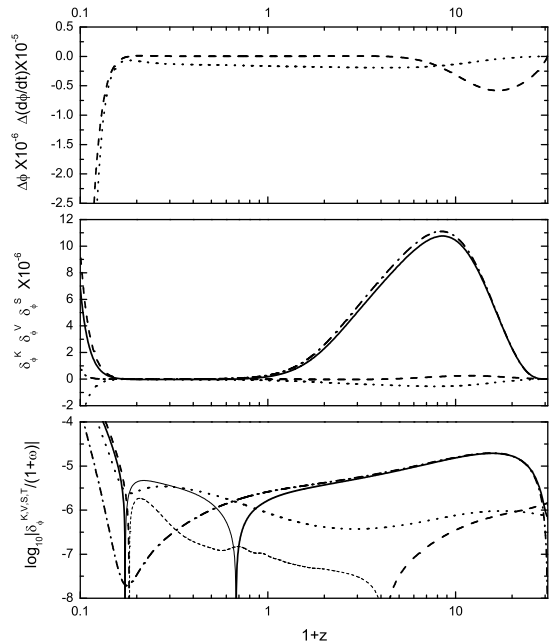}
\caption{\label{fig:sfsfdot2}Similar plot as in
Fig.~\ref{fig:sfsfdot1}, but for $r=0.25$. Upper panel: The
evolution of $\Delta \phi$ (dotted) and $\Delta \dot\phi$
(dashed); Middle panel: The evolution of $\delta_{\phi} ^K \times
10^{-6}$ (dashed) and $\delta_{\phi} ^V \times 10^{-6}$ (dotted)
and deformation energy density $\delta_{\phi}^S \times 10^{-6}$
(dash-dotted) and $(\delta_{\phi} ^K +\delta_{\phi} ^S+
\delta_{\phi} ^V) \times 10^{-6}$ (solid); Lower panel: The
evolution of $log_{10}|\delta_{\phi} ^K/(1+\omega)|$ (dashed) and
$log_{10}|\delta_{\phi} ^V/(1+\omega)|$ (dotted) and
$log_{10}|\delta_{\phi} ^S/(1+\omega)|$ (dash-dotted) and
$log_{10}|(\delta_{\phi} ^K + \delta_{\phi} ^S+\delta_{\phi}
^V)/(1+\omega)|$ (solid).}
\end{figure}

For shells between the center and the outer boundary, the
spatial variation of the field $\phi\prime$ contributes
significantly to the dark energy perturbations. Because of the
homogeneity of the background, the spatial perturbation term
always contributes positively to dark energy fluctuations. In
Fig.~\ref{fig:sfsfdot2}, we show different parts of dark energy
perturbations for the shell with $r=0.25$. The upper panel shows
the perturbations of $\phi$ (dotted line) and $\dot \phi$ (dashed
line), respectively. In comparison with those of $r=0$ (the upper
panel of Fig.~\ref{fig:sfsfdot1}), these perturbations are
smaller. The middle panel shows the potential (dotted line),
kinetic (dashed line), spatial (dash-dotted) terms, and the total
(solid line) of dark energy perturbations. For this shell, the
spatial term is the major perturbation term over a large redshift
range with $0<z<20$. The lower panel shows the corresponding
quantities over $(1+\omega)$ in log-scale. It is seen that for
$0<z<20$, the spatial term is much larger than the other two
terms, and can be as large as about $10^{-5}$. It is also noted
that $\phi\prime$ and the perturbation of $\dot \phi$ are
correlated with each other.

From Fig.~\ref{fig:space}, we see that the peak perturbation
occurs at $r \sim 0.1$ at high redshifts (solid and dashed lines).
It gradually moves outward, and eventually reaches $r \sim 0.2$.
We find that the peak position is in good accordance with the
shell within which the virialization happens. Such a shell
corresponds to the position of maximum $\phi \prime$.

   \begin{figure}
    \includegraphics[width=0.45\textwidth]{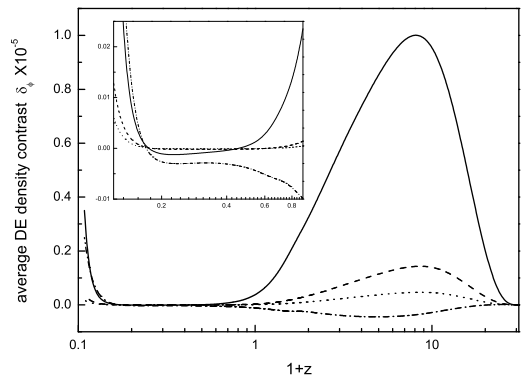}
    \caption{\label{fig:ave}The average $\bar \delta_{\phi} \times 10^{-5}.$
    The solid, dashed, dotted and dash-dotted lines are for the results averaging
    within $r=0.25, 0.5, 0.75 $ and $r=1$, respectively.}
    \end{figure}

Fig.~\ref{fig:ave} shows the average dark energy density
perturbations within $r=0.25$ (solid line), $0.5$ (dashed line),
$0.75$ (dotted line) and $1$ (dash-dotted line), respectively.
Firstly, we see the notable high peak at $z\sim 10$ for the solid
curve, which is contributed by the spatial derivative term
$\phi\prime$. As we average over larger regions outward, the dark
energy void balances its concentration in the inner region and
results smaller amplitude of dark energy perturbations. The
inserted region of the plot shows the zoom-in pattern of the
corresponding perturbations in very late times. The spatial
variation of the dark energy perturbations can be relevant in
considering the ISW effect in a collapsing overdense region, for which
we will investigate further in our future studies.

\begin{figure}
\includegraphics[width=0.45\textwidth]{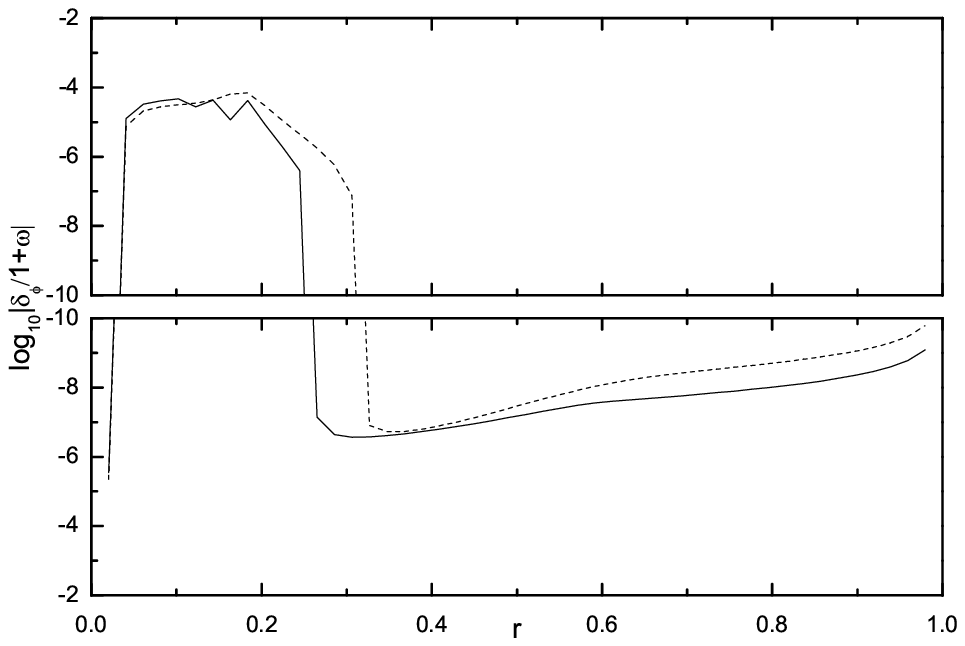}
\caption{\label{fig:tophat} The spatial distribution of $log_{10}|\delta
_{\phi}/(1+\omega)|$ at $z=0$. The solid and dashed lines are
for the compensated top-hat and the compensated Gaussian initial density profiles,
respectively.}
\end{figure}

\begin{figure}
\includegraphics[width=0.45\textwidth]{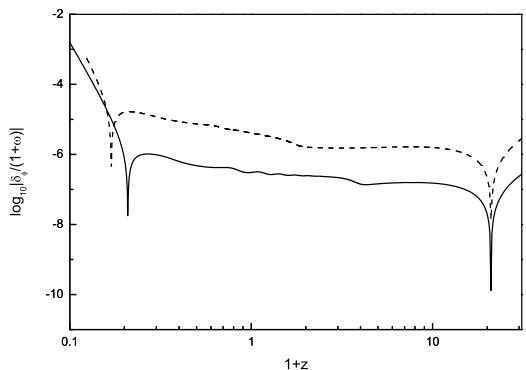}
\caption{\label{fig:gc} The evolution of $log_{10}|\delta
_{\phi}/(1+\omega)|$ for $r=0$. The solid line and dashed line are
for the galactic halo and cluster halo, respectively.}
\end{figure}

To see the dependence of our results on the chosen initial density
profile, we also perform the analyses with a compensated top-hat initial density distribution.
The mass contained in the overdense region and the central density perturbation
are set to be the same as those of the compensated Gaussian profile of Eq. (\ref{eq:delta}).
Thus in the top-hat profile, the overdense region is narrower than that of the
Gaussian one. Fig.~\ref{fig:tophat} shows the spatial distribution of
$\log_{10}|\delta _{\phi}/(1+\omega)|$ at $z=0$ with the solid and the dashed lines
for the top-hat and the Gaussian profiles, respectively.
It is seen that the overall behaviors of the two are similar, with positive dark energy
perturbations in the inner regions and negative ones in the outer regions. The narrower
range of the positive perturbations in the top-hat case is closely
related to its narrower matter overdense region than that of the Gaussian case.
Since the profile of the initial matter density perturbations affects the detailed collapsing
process, the specifics of dark energy perturbations are expected to depend on the initial
matter density profile. However, the overall level of the dark energy perturbations
is anticipated to be most sensitive to the total mass contained in the overdense region.
This can be seen from the following discussions. On the other hand, it is worthwhile to
investigate in detail the dependence of the dark energy perturbations on the initial mass density
profile, another direction that we can pursue in the future.

We now compare dark energy perturbations induced by the formation
of different sized halos. Fig.~\ref{fig:gc} shows the
dark energy perturbations at $r=0$ of the
galaxy-sized halo with $M=10^{12}\hbox{ M}_{\odot}$ (solid
line) and the cluster-sized halo with $M=10^{14}\hbox{
M}_{\odot}$ (dashed line). It is seen that the first transition
from positive to negative dark energy perturbations occur at
$z\sim 20$ for both cases. As we discussed previously, this
transition reflects the transition of the field from the fast
evolutionary stage with a steep potential to the slow evolutionary stage
with a relatively flat potential. Thus it is mainly determined by
the behavior of the field potential itself, which explains the
similarity seen in the two cases. On the other hand, the second
transition from negative to positive perturbations in late times
is closely related to the formation of the halos. It happens
earlier for galaxies than that of clusters. This is because of the higher density
perturbations initially set for the galaxy-sized halo, which leads to its
earlier formation than that of the cluster. For the overall
amplitude of dark energy perturbations, it is about an order of
magnitude larger for the cluster case than that of the
galaxy case. Thus it is mainly the mass of the system
rather than the initial amplitude of the dark matter density
perturbation that determines the overall amplitude of the
quintessence dark energy perturbation. This is consistent with the
expectation that quintessence dark energy perturbations show their
effects mainly on very large scales.

\section{Discussion and Conclusion}
In this paper, we analyze the quintessence dark energy
perturbations induced by the spherical collapse of dark matter
halos. We develop an iterative algorithm which allows us to solve
the field perturbations consistently from linear to quasi-linear
and to nonlinear stages of the halo formation. We find that the
dark energy perturbations depend sensitively on the field
potential, and thus on the background equation of state. In the
linear regime of the halo formation, the slower expansion near the
overdense region leads to the local evolution of the field to be
ahead of the background field. When the quintessence field is in
its fast evolutionary stage, $|\dot \phi|$ is accelerating
quickly. The perturbation of the kinetic dark energy term is
positive, and dominates over the negative perturbation of the
potential energy term. Adding in the positive perturbation from
the spatial term $\propto \phi'^2$ results a dark energy
enhancement everywhere around an overdense region at the very
early stage of the evolution. As the field enters the region where
the potential gets flatter, the $|\Delta \dot \phi|$ term is
smaller, and the potential dark energy perturbation takes over.
This gives rise to a negative perturbation, namely, dark energy
void, in central and outer regions where the spatial variation
term $\phi\prime$ is small in accordance with our chosen boundary
conditions. On the other hand, for regions in between, the
$\phi\prime$ term is relatively large and there are strong dark
energy concentrations. In the late stage when the considered
region collapses and virializes, the friction term $(\dot
R^\prime/R^\prime+2\dot R/R)\dot \phi$ changes sign to become a
driving source for the field evolution. Thus the time evolution of
$\phi$ is accelerated although the potential is nearly flat. This
leads to the domination of the positive kinetic dark energy
perturbation again in very late stages. The overall dark energy
perturbation is at the level $\sim 10^{-5}$ for cluster-sized
halos and $\sim 10^{-6}$ for galactic halos.

The work by \cite{DM07} is the earliest one to analyze the
dynamical dark energy evolution in the framework of spherical
collapse of dark matter halos. They employ a linearized numerical
approach, and thus only consider the linear stage of the halo
formation. They find a scalar field void associate with the
overdense region, and further argue that it should be deepened in
the nonlinear stage. While our results from the linear stage of
the halo formation are in qualitative agreement with theirs, some
differences are worth mentioning. First, in their calculation of
the dark energy perturbations, the term associated with the
spatial variation of the field $\phi\prime$ is not included. It is
seen from our analyses that in some spatial regions, this term can contribute significantly
to positive dark energy perturbations to form a dark energy concentration rather than
a void. Secondly, our results
(e.g., Fig.~\ref{fig:sfsfdot1}) show that the field void is indeed deepened
initially as expected, but then gradually becomes shallower. This
behavior can be understood by noting that in the regions where the
potential of the field is relatively flat, the velocity of the
field $|\dot \phi|$ decreases. As a result, the background field
gradually catches up the field around the overdense region, and
$|\Delta \phi|$ decreases. Therefore the depth that the field void
can go depends sensitively on the dark energy potential, and thus
the background equation of state.

Our study is mostly related to the study in \cite{MSS08}. In their
analyses, they adopt the method of matched asymptotic expansions
$\left( MAEs \right)$, in which different expansion parameters are
used at different regions, mainly $HR<<1$ in the interior region
and $\delta_m<<1$ in the exterior region. The two are required to
match with each other in the intermediate region. With such
approximations, they derive a relation between the dark energy
perturbations and the peculiar velocity $\delta v$ and the average
matter density contrast $\bar \delta_m$. Different treatments for
the linear, quasi-linear and nonlinear evolution of the halo
formation are then applied to obtain the information of $\delta v$
and $\bar \delta_m$, which finally give rise to the dark energy
perturbations. On the other hand, we employ a full numerical
treatment. The metric evolution is calculated in the Tolman
approximation with the effect of dark energy perturbations
neglected. We introduce an analytical description to mimic the
virialization of dark matter halos, which allows to carry out the
metric calculation all the way from linear to nonlinear stages of
the halo formation consistently. With the obtained metric, an
iterative algorithm is used to compute the dark energy
perturbations. Given different approaches, our results are in good
accordance with those of Mota et. al.  \cite{MSS08}. Both have
seen the transition for the dark energy perturbation from positive
to negative and to positive again. However, some quantitative
differences exist. Considering the central region, the second
transition from negative to positive dark energy perturbations
happens later from our results. We notice that in Mota et. al.
\cite{MSS08} the metric calculation in nonlinear stages is done by
ignoring the dark energy component completely. In our analyses,
the dark energy perturbations are neglected, but the background
dark energy component is kept in computing the metric evolution.

\begin{figure}
\includegraphics[width=0.45\textwidth]{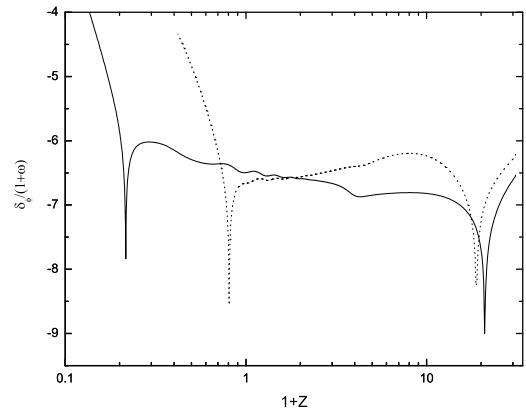}
\caption{\label{fig:mota}The influence of background dark energy
on dark energy perturbation. The solid line and the dashed line
are for the results with and without taking into account the
background dark energy in metric calculation.}
\end{figure}

For a test, we have done a calculation on metric by dropping the
dark energy component. The result on dark energy perturbations is
shown in Fig.~\ref{fig:mota} for the central region (dashed line)
together with the result with background dark energy component
included in metric calculations (solid line). It is seen that the
second transition indeed occurs earlier for the dashed line. This
basically reflects that the existence of background dark energy
affects the epoch of halo formation, which in turn affects the
behavior of dark energy perturbations. We also find that while
both catch the contribution of the $\phi\prime$ term to dark
energy perturbations, we obtain a larger amplitude of the term
than that of Mota et. al. \cite{MSS08}. This is closely related to
the different initial density profiles as well as the different
descriptions of the virialization process in the two studies.
This, to certain extents, indicates that dark energy perturbations
are sensitive to the details of their perturbing matter
distributions.

Our algorithm can be readily applied to study dark energy
perturbations of different models. Further investigations on their
cosmological implications and seeking for possible observational
effects of dark energy perturbations, e.g., the ISW effects around
superclusters and clusters of galaxies, are highly desired and
will be carried out as our future tasks.

\begin{acknowledgments}
We are very grateful to the referee for the encouraging comments and suggestions.
We thank Xinmin Zhang, Hong Li, Junqing Xia, and Mingzhe Li for
their helpful discussions. This research is supported in part by
the NSFC of China under grants 10373001, 10533010 and 10773001,
and the 973 program No.2007CB815401.

\end{acknowledgments}

\bibliography{paper}

\begin{thebibliography}{37}
\expandafter\ifx\csname natexlab\endcsname\relax\def\natexlab#1{#1}\fi
\expandafter\ifx\csname bibnamefont\endcsname\relax
  \def\bibnamefont#1{#1}\fi
\expandafter\ifx\csname bibfnamefont\endcsname\relax
  \def\bibfnamefont#1{#1}\fi
\expandafter\ifx\csname citenamefont\endcsname\relax
  \def\citenamefont#1{#1}\fi
\expandafter\ifx\csname url\endcsname\relax
  \def\url#1{\texttt{#1}}\fi
\expandafter\ifx\csname urlprefix\endcsname\relax\def\urlprefix{URL }\fi
\providecommand{\bibinfo}[2]{#2}
\providecommand{\eprint}[2][]{\url{#2}}

\bibitem[{\citenamefont{Astier et~al.}(2006)\citenamefont{Astier, Guy,
  Regnault, Pain, Aubourg, Balam, Basa, Carlberg, Fabbro, Fouchez
  et~al.}}]{Astier06}
\bibinfo{author}{\bibfnamefont{P.}~\bibnamefont{Astier}},
  \bibinfo{author}{\bibfnamefont{J.}~\bibnamefont{Guy}},
  \bibinfo{author}{\bibfnamefont{N.}~\bibnamefont{Regnault}},
  \bibinfo{author}{\bibfnamefont{R.}~\bibnamefont{Pain}},
  \bibinfo{author}{\bibfnamefont{E.}~\bibnamefont{Aubourg}},
  \bibinfo{author}{\bibfnamefont{D.}~\bibnamefont{Balam}},
  \bibinfo{author}{\bibfnamefont{S.}~\bibnamefont{Basa}},
  \bibinfo{author}{\bibfnamefont{R.~G.} \bibnamefont{Carlberg}},
  \bibinfo{author}{\bibfnamefont{S.}~\bibnamefont{Fabbro}},
  \bibinfo{author}{\bibfnamefont{D.}~\bibnamefont{Fouchez}},
  \bibnamefont{et~al.}, \bibinfo{journal}{Astron. Astrophys.}
  \textbf{\bibinfo{volume}{447}}, \bibinfo{pages}{31} (\bibinfo{year}{2006}).

\bibitem[{\citenamefont{Riess et~al.}(2007)\citenamefont{Riess, Strolger,
  Casertano, Ferguson, Mobasher, Gold, Challis, Filippenko, Jha, Li
  et~al.}}]{Riess07}
\bibinfo{author}{\bibfnamefont{A.~G.} \bibnamefont{Riess}},
  \bibinfo{author}{\bibfnamefont{L.-G.} \bibnamefont{Strolger}},
  \bibinfo{author}{\bibfnamefont{S.}~\bibnamefont{Casertano}},
  \bibinfo{author}{\bibfnamefont{H.~C.} \bibnamefont{Ferguson}},
  \bibinfo{author}{\bibfnamefont{B.}~\bibnamefont{Mobasher}},
  \bibinfo{author}{\bibfnamefont{B.}~\bibnamefont{Gold}},
  \bibinfo{author}{\bibfnamefont{P.~J.} \bibnamefont{Challis}},
  \bibinfo{author}{\bibfnamefont{A.~V.} \bibnamefont{Filippenko}},
  \bibinfo{author}{\bibfnamefont{S.}~\bibnamefont{Jha}},
  \bibinfo{author}{\bibfnamefont{W.}~\bibnamefont{Li}}, \bibnamefont{et~al.},
  \bibinfo{journal}{Astrophys.\ J.} \textbf{\bibinfo{volume}{659}},
  \bibinfo{pages}{98R} (\bibinfo{year}{2007}).

\bibitem[{\citenamefont{Copeland et~al.}(2006)\citenamefont{Copeland, Sami, and
  Tsujikawa}}]{Copeland06}
\bibinfo{author}{\bibfnamefont{E.~J.} \bibnamefont{Copeland}},
  \bibinfo{author}{\bibfnamefont{M.}~\bibnamefont{Sami}}, \bibnamefont{and}
  \bibinfo{author}{\bibfnamefont{S.}~\bibnamefont{Tsujikawa}},
  \bibinfo{journal}{Int. J. Mod. Phys. D} \textbf{\bibinfo{volume}{15}},
  \bibinfo{pages}{1753} (\bibinfo{year}{2006}).

\bibitem[{\citenamefont{Frieman et~al.}(2008)\citenamefont{Frieman, Turner, and
  Huterer}}]{Frieman08}
\bibinfo{author}{\bibfnamefont{J.}~\bibnamefont{Frieman}},
  \bibinfo{author}{\bibfnamefont{M.}~\bibnamefont{Turner}}, \bibnamefont{and}
  \bibinfo{author}{\bibfnamefont{D.}~\bibnamefont{Huterer}},
  \bibinfo{journal}{Annual Review of Astronomy and Astrophysics}
  \textbf{\bibinfo{volume}{46}}, \bibinfo{pages}{485} (\bibinfo{year}{2008}).

\bibitem[{\citenamefont{Linder}(2008)}]{Linder08}
\bibinfo{author}{\bibfnamefont{E.~V.} \bibnamefont{Linder}},
  \bibinfo{journal}{Gen. Relativ. Gravit.} \textbf{\bibinfo{volume}{40}},
  \bibinfo{pages}{329} (\bibinfo{year}{2008}).

\bibitem[{\citenamefont{Komatsu et~al.}(2009)\citenamefont{Komatsu, Dunkley,
  Nolta, Bennett, Gold, Hinshaw, Jarosik, Larson, Limon, Page
  et~al.}}]{Komatsu09}
\bibinfo{author}{\bibfnamefont{E.}~\bibnamefont{Komatsu}},
  \bibinfo{author}{\bibfnamefont{J.}~\bibnamefont{Dunkley}},
  \bibinfo{author}{\bibfnamefont{M.~R.} \bibnamefont{Nolta}},
  \bibinfo{author}{\bibfnamefont{C.~L.} \bibnamefont{Bennett}},
  \bibinfo{author}{\bibfnamefont{B.}~\bibnamefont{Gold}},
  \bibinfo{author}{\bibfnamefont{G.}~\bibnamefont{Hinshaw}},
  \bibinfo{author}{\bibnamefont{Jarosik}},
  \bibinfo{author}{\bibfnamefont{D.}~\bibnamefont{Larson}},
  \bibinfo{author}{\bibfnamefont{M.}~\bibnamefont{Limon}},
  \bibinfo{author}{\bibfnamefont{L.}~\bibnamefont{Page}}, \bibnamefont{et~al.},
  \bibinfo{journal}{Astrophys. J. S.} \textbf{\bibinfo{volume}{180}},
  \bibinfo{pages}{330} (\bibinfo{year}{2009}).

\bibitem[{\citenamefont{Li et~al.}(2008{\natexlab{a}})\citenamefont{Li, Xia,
  Fan, and Zhang}}]{Li08JCAP}
\bibinfo{author}{\bibfnamefont{H.}~\bibnamefont{Li}},
  \bibinfo{author}{\bibfnamefont{J.-Q.} \bibnamefont{Xia}},
  \bibinfo{author}{\bibfnamefont{Z.}~\bibnamefont{Fan}}, \bibnamefont{and}
  \bibinfo{author}{\bibfnamefont{X.}~\bibnamefont{Zhang}},
  \bibinfo{journal}{J.\ Cosmol.\ Astropart.\ Phys.}
  \textbf{\bibinfo{volume}{10}}, \bibinfo{pages}{046}
  (\bibinfo{year}{2008}{\natexlab{a}}).

\bibitem[{\citenamefont{Klypin et~al.}(2003)\citenamefont{Klypin, Macci\`o,
  Mainini, and Bonometto}}]{Klypin03}
\bibinfo{author}{\bibfnamefont{A.}~\bibnamefont{Klypin}},
  \bibinfo{author}{\bibfnamefont{A.~V.} \bibnamefont{Macci\`o}},
  \bibinfo{author}{\bibfnamefont{R.}~\bibnamefont{Mainini}}, \bibnamefont{and}
  \bibinfo{author}{\bibfnamefont{A.}~\bibnamefont{Bonometto}},
  \bibinfo{journal}{Astrophys. J.} \textbf{\bibinfo{volume}{599}},
  \bibinfo{pages}{31} (\bibinfo{year}{2003}).

\bibitem[{\citenamefont{Macci\`o}(2005)}]{Maccio05}
\bibinfo{author}{\bibfnamefont{A.~V.} \bibnamefont{Macci\`o}},
  \bibinfo{journal}{Mon.\ Not.\ Roy.\ Astron.\ Soc.}
  \textbf{\bibinfo{volume}{361}}, \bibinfo{pages}{1256} (\bibinfo{year}{2005}).

\bibitem[{\citenamefont{LeDelliou}(2006)}]{LeDelliou06}
\bibinfo{author}{\bibfnamefont{M.}~\bibnamefont{LeDelliou}},
  \bibinfo{journal}{J.\ Cosmol.\ Astropart.\ Phys.}
  \textbf{\bibinfo{volume}{01}}, \bibinfo{pages}{021} (\bibinfo{year}{2006}).

\bibitem[{\citenamefont{Mainini and Bonometto}(2007)}]{Mainini07}
\bibinfo{author}{\bibfnamefont{R.}~\bibnamefont{Mainini}} \bibnamefont{and}
  \bibinfo{author}{\bibfnamefont{S.}~\bibnamefont{Bonometto}},
  \bibinfo{journal}{J.\ Cosmol.\ Astropart.\ Phys.}
  \textbf{\bibinfo{volume}{09}}, \bibinfo{pages}{017} (\bibinfo{year}{2007}).

\bibitem[{\citenamefont{Unnikrishnan et~al.}(2008)\citenamefont{Unnikrishnan,
  Jassal, and Seshadri}}]{Unnikr08}
\bibinfo{author}{\bibfnamefont{S.}~\bibnamefont{Unnikrishnan}},
  \bibinfo{author}{\bibfnamefont{H.~K.} \bibnamefont{Jassal}},
  \bibnamefont{and} \bibinfo{author}{\bibfnamefont{T.~R.}
  \bibnamefont{Seshadri}}, \bibinfo{journal}{Phys.\ Rev.\ D.}
  \textbf{\bibinfo{volume}{78}}, \bibinfo{pages}{123504}
  (\bibinfo{year}{2008}).

\bibitem[{\citenamefont{Amendola}(2004)}]{Amendola04}
\bibinfo{author}{\bibfnamefont{L.}~\bibnamefont{Amendola}},
  \bibinfo{journal}{Phys.\ Rev.\ D.} \textbf{\bibinfo{volume}{69}},
  \bibinfo{pages}{103524} (\bibinfo{year}{2004}).

\bibitem[{\citenamefont{Zhao et~al.}(2005)\citenamefont{Zhao, Xia, Li, Feng,
  and Zhang}}]{Zhao05}
\bibinfo{author}{\bibfnamefont{G.-B.} \bibnamefont{Zhao}},
  \bibinfo{author}{\bibfnamefont{J.-Q.} \bibnamefont{Xia}},
  \bibinfo{author}{\bibfnamefont{M.}~\bibnamefont{Li}},
  \bibinfo{author}{\bibfnamefont{B.}~\bibnamefont{Feng}}, \bibnamefont{and}
  \bibinfo{author}{\bibfnamefont{X.}~\bibnamefont{Zhang}},
  \bibinfo{journal}{Phys.\ Rev.\ D.} \textbf{\bibinfo{volume}{72}},
  \bibinfo{pages}{123515} (\bibinfo{year}{2005}).

\bibitem[{\citenamefont{Zhao et~al.}(2007)\citenamefont{Zhao, Xia, Feng, and
  Zhang}}]{Zhao07}
\bibinfo{author}{\bibfnamefont{G.-B.} \bibnamefont{Zhao}},
  \bibinfo{author}{\bibfnamefont{J.-Q.} \bibnamefont{Xia}},
  \bibinfo{author}{\bibfnamefont{B.}~\bibnamefont{Feng}}, \bibnamefont{and}
  \bibinfo{author}{\bibfnamefont{X.}~\bibnamefont{Zhang}},
  \bibinfo{journal}{Int. J. Mod. Phys. D} \textbf{\bibinfo{volume}{16}},
  \bibinfo{pages}{1229} (\bibinfo{year}{2007}).

\bibitem[{\citenamefont{Li et~al.}(2008{\natexlab{b}})\citenamefont{Li, Xia,
  Zhao, Fan, and Zhan}}]{Li08ApJ}
\bibinfo{author}{\bibfnamefont{H.}~\bibnamefont{Li}},
  \bibinfo{author}{\bibfnamefont{J.-Q.} \bibnamefont{Xia}},
  \bibinfo{author}{\bibfnamefont{G.-B.} \bibnamefont{Zhao}},
  \bibinfo{author}{\bibfnamefont{Z.-H.} \bibnamefont{Fan}}, \bibnamefont{and}
  \bibinfo{author}{\bibfnamefont{X.}~\bibnamefont{Zhan}},
  \bibinfo{journal}{Astrophys.\ J.} \textbf{\bibinfo{volume}{683}},
  \bibinfo{pages}{L1} (\bibinfo{year}{2008}{\natexlab{b}}).

\bibitem[{\citenamefont{LaVacca and Colombo}(2008)}]{LaVacca08}
\bibinfo{author}{\bibfnamefont{G.}~\bibnamefont{LaVacca}} \bibnamefont{and}
  \bibinfo{author}{\bibfnamefont{L.~P.~L.} \bibnamefont{Colombo}},
  \bibinfo{journal}{J.\ Cosmol.\ Astropart.\ Phys.}
  \textbf{\bibinfo{volume}{04}}, \bibinfo{pages}{007} (\bibinfo{year}{2008}).

\bibitem[{\citenamefont{Granett et~al.}(2008)\citenamefont{Granett, Neyrinck,
  and Szapudi}}]{Granett08}
\bibinfo{author}{\bibfnamefont{B.~R.} \bibnamefont{Granett}},
  \bibinfo{author}{\bibfnamefont{M.~C.} \bibnamefont{Neyrinck}},
  \bibnamefont{and} \bibinfo{author}{\bibfnamefont{I.}~\bibnamefont{Szapudi}},
  \bibinfo{journal}{Astrophys.\ J.} \textbf{\bibinfo{volume}{683}},
  \bibinfo{pages}{99} (\bibinfo{year}{2008}).

\bibitem[{\citenamefont{Sakai and Inoue}(2008)}]{Sakai08}
\bibinfo{author}{\bibfnamefont{N.}~\bibnamefont{Sakai}} \bibnamefont{and}
  \bibinfo{author}{\bibfnamefont{K.~T.} \bibnamefont{Inoue}},
  \bibinfo{journal}{Phys.\ Rev.\ D.} \textbf{\bibinfo{volume}{78}},
  \bibinfo{pages}{063510} (\bibinfo{year}{2008}).

\bibitem[{\citenamefont{Masina and Notari}(2009)}]{Masina09}
\bibinfo{author}{\bibfnamefont{I.}~\bibnamefont{Masina}} \bibnamefont{and}
  \bibinfo{author}{\bibfnamefont{A.}~\bibnamefont{Notari}},
  \bibinfo{journal}{J.\ Cosmol.\ Astropart.\ Phys.}
  \textbf{\bibinfo{volume}{02}}, \bibinfo{pages}{019} (\bibinfo{year}{2009}).

\bibitem[{\citenamefont{Wang and Steinhardt}(1998)}]{WS98}
\bibinfo{author}{\bibfnamefont{L.~M.} \bibnamefont{Wang}} \bibnamefont{and}
  \bibinfo{author}{\bibfnamefont{P.~J.} \bibnamefont{Steinhardt}},
  \bibinfo{journal}{Astrophys.\ J.} \textbf{\bibinfo{volume}{508}},
  \bibinfo{pages}{483} (\bibinfo{year}{1998}).

\bibitem[{\citenamefont{Maor and Lahav}(2005)}]{ML04}
\bibinfo{author}{\bibfnamefont{I.}~\bibnamefont{Maor}} \bibnamefont{and}
  \bibinfo{author}{\bibfnamefont{O.}~\bibnamefont{Lahav}},
  \bibinfo{journal}{JCAP} \textbf{\bibinfo{volume}{0507}}, \bibinfo{pages}{003}
  (\bibinfo{year}{2005}).

\bibitem[{\citenamefont{Mota and van~de Bruck}(2004)}]{MB04}
\bibinfo{author}{\bibfnamefont{D.~F.} \bibnamefont{Mota}} \bibnamefont{and}
  \bibinfo{author}{\bibfnamefont{C.}~\bibnamefont{van~de Bruck}},
  \bibinfo{journal}{Astron. Astrophys.} \textbf{\bibinfo{volume}{421}},
  \bibinfo{pages}{71} (\bibinfo{year}{2004}).

\bibitem[{\citenamefont{Nunes and Mota}(2006)}]{NM06}
\bibinfo{author}{\bibfnamefont{N.~J.} \bibnamefont{Nunes}} \bibnamefont{and}
  \bibinfo{author}{\bibfnamefont{D.~F.} \bibnamefont{Mota}},
  \bibinfo{journal}{Mon.\ Not.\ Roy.\ Astron.\ Soc.}
  \textbf{\bibinfo{volume}{368}}, \bibinfo{pages}{751} (\bibinfo{year}{2006}).

\bibitem[{\citenamefont{Langlois and Vernizzi}(2007)}]{Langlois07}
\bibinfo{author}{\bibfnamefont{D.}~\bibnamefont{Langlois}} \bibnamefont{and}
  \bibinfo{author}{\bibfnamefont{F.}~\bibnamefont{Vernizzi}},
  \bibinfo{journal}{J.\ Cosmol.\ Astropart.\ Phys.}
  \textbf{\bibinfo{volume}{02}}, \bibinfo{pages}{017} (\bibinfo{year}{2007}).

\bibitem[{\citenamefont{Abramo et~al.}(2007)\citenamefont{Abramo, Batista,
  Liberato, and Rosenfeld}}]{Abramo07}
\bibinfo{author}{\bibfnamefont{L.~R.} \bibnamefont{Abramo}},
  \bibinfo{author}{\bibfnamefont{R.~C.} \bibnamefont{Batista}},
  \bibinfo{author}{\bibfnamefont{L.}~\bibnamefont{Liberato}}, \bibnamefont{and}
  \bibinfo{author}{\bibfnamefont{R.}~\bibnamefont{Rosenfeld}},
  \bibinfo{journal}{J.\ Cosmol.\ Astropart.\ Phys.}
  \textbf{\bibinfo{volume}{11}}, \bibinfo{pages}{012} (\bibinfo{year}{2007}).

\bibitem[{\citenamefont{Mota}(2008)}]{MOTA08JCAP}
\bibinfo{author}{\bibfnamefont{D.~F.} \bibnamefont{Mota}},
  \bibinfo{journal}{J.\ Cosmol.\ Astropart.\ Phys.}
  \textbf{\bibinfo{volume}{09}}, \bibinfo{pages}{006} (\bibinfo{year}{2008}).

\bibitem[{\citenamefont{Dutta and Maor}(2007)}]{DM07}
\bibinfo{author}{\bibfnamefont{S.}~\bibnamefont{Dutta}} \bibnamefont{and}
  \bibinfo{author}{\bibfnamefont{I.}~\bibnamefont{Maor}},
  \bibinfo{journal}{Phys.\ Rev.\ D} \textbf{\bibinfo{volume}{75}},
  \bibinfo{pages}{063507} (\bibinfo{year}{2007}).

\bibitem[{\citenamefont{Mota et~al.}(2008)\citenamefont{Mota, Shaw, and
  Silk}}]{MSS08}
\bibinfo{author}{\bibfnamefont{D.~F.} \bibnamefont{Mota}},
  \bibinfo{author}{\bibfnamefont{D.~J.} \bibnamefont{Shaw}}, \bibnamefont{and}
  \bibinfo{author}{\bibfnamefont{J.}~\bibnamefont{Silk}},
  \bibinfo{journal}{Astrophys.\ J.} \textbf{\bibinfo{volume}{675}},
  \bibinfo{pages}{29M} (\bibinfo{year}{2008}).

\bibitem[{\citenamefont{Barreiro et~al.}(2000)\citenamefont{Barreiro, Copeland,
  and N.J.Nunes}}]{Barreiro00}
\bibinfo{author}{\bibfnamefont{T.}~\bibnamefont{Barreiro}},
  \bibinfo{author}{\bibfnamefont{E.~J.} \bibnamefont{Copeland}},
  \bibnamefont{and} \bibinfo{author}{\bibnamefont{N.J.Nunes}},
  \bibinfo{journal}{Phys.\ Rev.\ D} \textbf{\bibinfo{volume}{61}},
  \bibinfo{pages}{127301} (\bibinfo{year}{2000}).

\bibitem[{\citenamefont{Press et~al.}(2002)\citenamefont{Press, Teukolsky,
  Vetterling, and Flannery}}]{NR02}
\bibinfo{author}{\bibfnamefont{W.~H.} \bibnamefont{Press}},
  \bibinfo{author}{\bibfnamefont{S.~A.} \bibnamefont{Teukolsky}},
  \bibinfo{author}{\bibfnamefont{W.~T.} \bibnamefont{Vetterling}},
  \bibnamefont{and} \bibinfo{author}{\bibfnamefont{B.~P.}
  \bibnamefont{Flannery}}, \emph{\bibinfo{title}{Numerical Recipes in C++}}
  (\bibinfo{publisher}{Cambridge University Press}, \bibinfo{year}{2002}).

\bibitem[{\citenamefont{Dunkley et~al.}(2009)\citenamefont{Dunkley, Komatsu,
  Nolta, Spergel, Larson, Hinshaw, Page, Bennett, Gold, Jarosik
  et~al.}}]{WMAP5}
\bibinfo{author}{\bibfnamefont{J.}~\bibnamefont{Dunkley}},
  \bibinfo{author}{\bibfnamefont{E.}~\bibnamefont{Komatsu}},
  \bibinfo{author}{\bibfnamefont{M.~R.} \bibnamefont{Nolta}},
  \bibinfo{author}{\bibfnamefont{D.~N.} \bibnamefont{Spergel}},
  \bibinfo{author}{\bibnamefont{Larson}},
  \bibinfo{author}{\bibfnamefont{G.}~\bibnamefont{Hinshaw}},
  \bibinfo{author}{\bibfnamefont{L.}~\bibnamefont{Page}},
  \bibinfo{author}{\bibfnamefont{C.~L.} \bibnamefont{Bennett}},
  \bibinfo{author}{\bibfnamefont{B.}~\bibnamefont{Gold}},
  \bibinfo{author}{\bibfnamefont{N.}~\bibnamefont{Jarosik}},
  \bibnamefont{et~al.}, \bibinfo{journal}{Astrophys\ J.\ S.}
  \textbf{\bibinfo{volume}{180}}, \bibinfo{pages}{306} (\bibinfo{year}{2009}).

\bibitem[{\citenamefont{Eisenstein and Hu}(1998)}]{EH98}
\bibinfo{author}{\bibfnamefont{D.~J.} \bibnamefont{Eisenstein}}
  \bibnamefont{and} \bibinfo{author}{\bibfnamefont{W.}~\bibnamefont{Hu}},
  \bibinfo{journal}{Astrophys.\ J.} \textbf{\bibinfo{volume}{496}},
  \bibinfo{pages}{605} (\bibinfo{year}{1998}).

\bibitem[{\citenamefont{Shaw and Mota}(2008)}]{SHAW08}
\bibinfo{author}{\bibfnamefont{D.~J.} \bibnamefont{Shaw}} \bibnamefont{and}
  \bibinfo{author}{\bibfnamefont{D.~F.} \bibnamefont{Mota}},
  \bibinfo{journal}{Astrophys.\ J. S.} \textbf{\bibinfo{volume}{174}},
  \bibinfo{pages}{277} (\bibinfo{year}{2008}).

\bibitem[{\citenamefont{Hamilton et~al.}(1991)\citenamefont{Hamilton, Kumar,
  Lu, and AlexMatthews}}]{HAMILTON91}
\bibinfo{author}{\bibfnamefont{A.~J.~S.} \bibnamefont{Hamilton}},
  \bibinfo{author}{\bibfnamefont{P.}~\bibnamefont{Kumar}},
  \bibinfo{author}{\bibfnamefont{E.}~\bibnamefont{Lu}}, \bibnamefont{and}
  \bibinfo{author}{\bibnamefont{AlexMatthews}}, \bibinfo{journal}{Astrophys.
  J.} \textbf{\bibinfo{volume}{374}}, \bibinfo{pages}{L1}
  (\bibinfo{year}{1991}).

\bibitem[{\citenamefont{Navarro et~al.}(1996)\citenamefont{Navarro, Frenk, and
  White}}]{NFW96}
\bibinfo{author}{\bibfnamefont{J.~F.} \bibnamefont{Navarro}},
  \bibinfo{author}{\bibfnamefont{C.~S.} \bibnamefont{Frenk}}, \bibnamefont{and}
  \bibinfo{author}{\bibfnamefont{S.~D.~M.} \bibnamefont{White}},
  \bibinfo{journal}{Astrophys.\ J.} \textbf{\bibinfo{volume}{462}},
  \bibinfo{pages}{563} (\bibinfo{year}{1996}).

\bibitem[{\citenamefont{Navarro et~al.}(1997)\citenamefont{Navarro, Frenk, and
  White}}]{NFW97}
\bibinfo{author}{\bibfnamefont{J.~F.} \bibnamefont{Navarro}},
  \bibinfo{author}{\bibfnamefont{C.~S.} \bibnamefont{Frenk}}, \bibnamefont{and}
  \bibinfo{author}{\bibfnamefont{S.~D.~M.} \bibnamefont{White}},
  \bibinfo{journal}{Astrophys.\ J.} \textbf{\bibinfo{volume}{490}},
  \bibinfo{pages}{493} (\bibinfo{year}{1997}).

\end{thebibliography}

\end{document}